\documentclass{aa} 

\usepackage{graphicx}

\newcommand{\bm}[1]{\mbox{{\boldmath $#1$}}}
\newcommand{\der}[3]{\frac{d^{#3} {#1}}{d {#2}^{#3}}}
\newcommand{\pder}[3]{\frac{{\partial}^{#3} {#1}}{{\partial} {#2}^{#3}}}

\newcommand{\sign}{{\rm sign}}
\newcommand{\ecr}{e_{\rm cr}}

\newcommand{\Myr}{{\rm \ Myr}}
\newcommand{\pc}{{\rm \ pc}}
\newcommand{\erg}{{\rm \ erg}}
\newcommand{\km}{{\rm \ km}}
\newcommand{\cm}{{\rm \ cm}}
\newcommand{\s}{{\rm \ s}}

\begin{document}

\title{Incorporation of cosmic ray transport into the ZEUS MHD code}
\subtitle{Application for studies of Parker instability in the ISM}

\author{M. Hanasz \inst{1}
\and
H. Lesch \inst{2}}

\authorrunning{Hanasz \& Lesch }
\titlerunning{Incorporation of cosmic ray transport into the ZEUS code}
\offprints{M. Hanasz}

\institute{
Toru\'n Centre for Astronomy, Nicholas Copernicus University
PL-87148 Piwnice/Toru\'n, Poland, \email{mhanasz@astri.uni.torun.pl}
\and
University Observatory, M\"unchen University, Scheinerstr. 1, D-81679, Germany and 
Center for Interdisciplinary Plasma Science (CIPS),
\email{lesch@usm.uni-muenchen.de}
}

\date{Received 16 June 2003/ Accepted 12 September 2003}
\setcounter{footnote}{0}

\abstract{ We present a numerical algorithm for the incorporation of the
active  cosmic ray transport, into the  ZEUS-3D magnetohydrodynamical code. 
The cosmic ray transport is described by the diffusion-advection equation. The
applied form of the diffusion tensor allows for anisotropic diffusion of cosmic
rays along and across the magnetic field direction, which is controlled by two
parameters: the parallel and perpendicular diffusion coefficients.
The implemented numerical algorithm is tested by comparison of the diffusive
transport of cosmic rays to analytical solutions of the diffusion equation. Our
method is  numerically  stable for a wide range of diffusion coefficients,
including the realistic values inferred from the observational data for the
Milky Way of about $ 6 \times 10^{28} \cm^2 \s^{-1}$. The presented algorithm is applied for  
exemplary simulations of the the Parker instability  triggered by cosmic rays injected 
by a single SN remnant. 
}
\maketitle

\section{Introduction}

One of the major components of the interstellar medium (ISM), the cosmic ray
(CR) gas, consists of relativistic electrons, protons and heavier atomic nuclei
(see eg. Berezinski et al. 1990). It was shown beyond any doubt that the cosmic
ray particles are  accelerated in the process of diffusive acceleration by
shocks associated with supernova remnants  (SNR) in galactic disks (e.g. Koyama
et al. 1995).  Recent models suggest that the conversion
rate of the supernova energy into cosmic ray energy is in the range of 10 - 50
\% (see eg. Jones 1998 and references therein). The total kinetic energy output
from a single supernova is of the order of  $10^{51} \erg$, therefore the total
CR energy per unit volume, produced within a supernova remnant, is significant
as compared to thermal, kinetic and magnetic energy densities of the ISM. 

Although the velocity of individual CR particles is close to the speed of
light, the bulk motion of CR is diffusive and the CR bulk speed is of the order
of Alfv\'en speed, i.e. typically a few tens of km/s. Recent studies by
Giacalone and Jokipii (1999) and Jokipii (1999) suggest that the diffusion of
cosmic ray gas  in a turbulent magnetic field proceeds preferentially along the
direction of the mean magnetic field.  In our case the term cosmic rays
means protons and nuclei but not  electrons, since their contribution to the
pressure is negligible. The estimations made by Strong and Moskalenko ( 1998)
based on the GALPROP model provide the parallel  diffusion coefficients of the
order of $K_\parallel = 6 \times 10^{28} \cm^2 \s^{-1}$. This value is 2-3
orders of magnitude larger than the diffusion coefficient for turbulent mixing
of the ISM. The large energies carried by the cosmic ray component as well as
its highly diffusive nature imply that the cosmic ray component cannot be
neglected in  the studies of dynamics of the ISM. That statement follows
directly from  investigations of stability of the ISM on spatial scales of the
order of one up to a few kiloparsecs. Parker (1966, 1967) found that the
multicomponent interstellar medium stratified by vertical gravity is subject to
an instability which is caused by the buoyancy of the  weightless ISM
components, i.e magnetic field and cosmic rays. 

The Parker instability has been extensively studied by numerous authors in the
linear approximation under various circumstances like different disk gravity
models (Giz \& Shu 1993; Kim \& Hong 1998), rigid and differential rotation
(Shu 1974; Foglizzo \& Tagger 1994, 1995; Hanasz \& Lesch 1997) the presence of
random magnetic field component (Parker \& Jokipii 2000; Kim \& Ryu 2001) and 
nonadiabatic effects in the ISM (Kosinski \& Hanasz 2003). 

Majority of the work was done within the limit of very
large diffusion of cosmic rays along magnetic field lines and negligible
diffusion across magnetic field lines. The effect of finite diffusion was
studied by Kuznetsov and Ptuskin (1983) and recently by Ryu et al. (2003), who
demonstrate that within the linear approximation incorporation of the
diffusion-advection equation and realistic diffusion coefficients leads to
results consistent with the mentioned simplified description of cosmic ray
transport. According to the analysis done by Ryu et al. (2003) the finiteness of
the diffusion coefficient decreases the growth rate of the Parker instability.

On the other hand numerical studies of the Parker instability  investigate 
the effects of uniform vertical gravity (Kim et al. 1998), realistic
vertical gravity (Kim et al. 2000), selfgravity (Chou et al. 2000), the
effects of spiral arms (Franco et al. 2002), finite resistivity (Hanasz et al.
2002, Tanuma 2003; Kowal et al 2003),  partial ionization (Birk 2002) and
coupling to other disk instabilities (Kim, Ostriker \& Stone  2002).  

Surprisingly, the powerful cosmic ray component, which according to the linear
analysis  is crucial for the growth rate of the Parker instability, is
neglected in numerical studies except the recent paper by Hanasz \& Lesch
(2000) who  incorporate the diffusion-advection equation for the cosmic ray
transport along magnetic field and study the Parker instability triggered by
cosmic ray injection in SN remnants, applying the thin fluxtube approximation. 
That paper demonstrates the importance of the cosmic ray component for the
global dynamics of the ISM, including the hydromagnetic dynamo effect.

In this paper we describe how to  introduce the cosmic ray component, within
the diffusion-advection equation, into the ZEUS-3D MHD code (Stone \&
Norman 1992a, 1992b) developped at the Laboratory of Computational Astrophysics
(NCSA, University of Illinois at Urbana Champaign, see {\em
http://lca.ncsa.uiuc.edu/lca\_codes\_docs.html}). The ZEUS code uses a time-explicit, operator-split,
finite-difference method to solve the MHD equations on a staggered mesh. The MHD
algorithm employs the constrained transport formalism and the method of
characteristics for accurate propagation of Alfv\'en waves (Evans \& Hawley 1988, 
Hawley \& Stone 1995).

In the present paper we focus on the numerical method for the active cosmic-ray
transport. In Section 2. we introduce the set of basic equations. The numerical
algorithm is described in section 3, followed by  tests of the numerical method
and a  comparison of some results of computations to analytical solutions in
section 4. Section 5 contains as an example an application of the extended code
for studies of the Parker instability  triggered by the cosmic ray injection in
a single SN remnant. Finally in Section 6. we summarize our results.

\section{Equations of MHD including cosmic-ray transport}

The diffusive cosmic ray (CR) transport on macroscopic astrophysical scales is
described by the diffusion-advection equation. Following Schlickeiser and
Lerche (1985)  we apply the following form of the transport equation 
\begin{equation}
\pder{e_{\rm cr}}{t}{} +\bm{\nabla }\left( e_{\rm cr} \bm{v}\right) 
= \bm{\nabla} \left(\hat{K} \bm{\nabla} e_{cr} \right) 
- p_{\rm cr} \left( \bm{\nabla} \cdot \bm{v} \right) 
+ Q, \label{diff-adv-eq} 
\end{equation} 
where $e_{\rm cr}$ and $p_{\rm cr}= (\gamma_{\rm cr} -1) e_{\rm cr} $ are the
cosmic ray energy density and cosmic ray pressure, $\gamma_{\rm cr}$ (=4/3 in
this paper) denotes the ratio of the specific heats of the relativistic cosmic
ray gas,  $\hat{K}$ presents the diffusion tensor,  $\bm{v}$ is velocity of the
thermal gas and $Q$ is the source term for the CR energy density resulting from
the cosmic ray injection by supernova remnants (SNR) or alternative sources. 
We note that eq. (\ref{diff-adv-eq}) assumes that cosmic rays are treated
as a magnetized relativistic  gas. This assumption holds as long as the
particles are tied to the magnetic  field, i.e. as long as they gyroradius is
significantly smaller than the  characteristic spatial scales of the magnetic
field. Only cosmic rays with  ultrahigh energies  are ruled out by our approach
since their gyroradius is  larger than the thickness of the galactic disk. Eq.
(\ref{diff-adv-eq}) implicitly assumes that the  diffusion tensor describes the
interaction of charged particles with magnetic fluctuations which appear on
spatial scales  considerably smaller than the characteristic scales in the
interstellar medium. In the limit  of low energies ($< 100$ MeV) the the
current approximation is not valid because cosmic-ray energy losses
become important, although, the cosmic ray pressure still resides in that
range. 

In the present approach we apply the concept of anisotropic cosmic ray
diffusion following Giaccalone and Jokipii (1999), Jokipii (1999), Hanasz and
Lesch (2000) and Ryu et al (2003). In order to describe formally the
anisotropic cosmic ray diffusion we implement the diffusion tensor (see e.g.
Ryu et al. 2003)  of the form

\begin{equation}
K_{ij} = K_{\rm \perp} \delta_{ij} + (K_\parallel - K_{\rm \perp}) n_i n_j,
\label{diftens}
\end{equation}
where $n_i = B_i/B$ are components of the unit vectors tangent to magnetic
field lines. 
The above cosmic-ray transport equation supplements the standard
set of ideal MHD equations

\begin{equation}\pder{\rho}{t}{}+ \bm{\nabla} \cdot (\rho \bm{v}) = 0, \label{eqofconti}
\end{equation}
\begin{equation}
\pder{e}{t}{} +\bm{\nabla}\cdot \left( e \bm{v}\right) = - p \left( \bm{\nabla} \cdot \bm{v} 
\right), \label{enereq} 
\end{equation} 
\begin{equation}
\pder{\bm{v}}{t}{} + (\bm{v} \cdot \bm{\nabla})\bm{v} = -\frac{1}{\rho} \bm{\nabla}
   \left(p + p_{\rm cr} + \frac{B^2}{8\pi}\right)  
   + \frac{\bm{B \cdot \nabla B}}{4 \pi \rho},
   \label{eqofmot}
\end{equation}
\begin{equation}
\pder{\bm{B}}{t}{} = \bm{\nabla} \times \left( \bm{v} \times \bm{B}
\right). \label{indeq}
\end{equation}
where the gradient of cosmic ray pressure $\bm{\nabla} p_{\rm cr}$ has been
included in the equation of gas motion (see Berezinski et al. 1990). The
other symbols have their usual meaning.

\section{Numerical algorithm for the cosmic ray transport}

The actual form of the diffusion-advection equation (\ref{diff-adv-eq}) is similar
to the energy equation (\ref{enereq}) except the diffusion term, therefore we
incorporate an integration algorithm for the advection part of the
cosmic-ray transport, following the method of integration of the energy
equation (see Stone and Norman 1992a,b).

The integration method for the energy equation consists of a source step and
a transport step. In the source step  the $-p \bm{\nabla} \bm{v}$ term is
evaluated together with possible explicit sources of the internal energy. In the
transport step  fluxes of internal energy, through cell boundaries,  are
computed using directional splitting. The total amount of internal energy
within each cell is subsequently updated according to the sum of fluxes through
all cell boundaries.

The implementation of the cosmic ray transport requires an additional contribution
of diffusive fluxes  
\begin{equation}
\bm{F}^d_{cr} = \hat{K} \bm{\nabla} e_{cr} \label{cr-flux}
\end{equation}
corresponding to the term $\bm{\nabla} \left( \hat{K} \bm{\nabla} e_{cr} \right)$ in the
cosmic ray diffusion-advection equation. The tensorial form of the diffusion
coefficient $\hat{K}$ is used to describe the anisotropic cosmic ray diffusion.

In order to incorporate the diffusion of cosmic rays in the numerical
algorithm, along the magnetic field lines one should compute first components
of the unit vector $\bm{n} = \bm{B}/B$ parallel to the magnetic field
direction, separately for each cell face. Since in the ZEUS code vector field
components are centered on different cell faces  an averaging is necessary for
these vector components which are parallel to the given cell face. For
instance,  the magnetic field components on 1-faces (assigned with the
superscript  '1f') are given by 
\begin{eqnarray}
\bm{B}^{\rm 1f}_{(i,j,k)}=[B_{1(i,  j,  k)},\makebox[5.5cm]{}\\
 0.25(B_{2(i,j,k)}+B_{2(i-1,j,k)}+B_{2(i-1,j+1,k)}+B_{2(i,j+1,k))},
 \nonumber \\
 0.25(B_{3(i,j,k)}+B_{3(i-1,j,k)}+B_{3(i-1,j,k+1)}+B_{3(i,  j,k+1)})].
 \nonumber 
\end{eqnarray}
The field components on the other faces are computed analogously.

The next step is a computation of cosmic-ray diffusive fluxes across cell
interfaces. This requires a prior computation of components of the gradient of
cosmic-ray energy density. All three gradient components contributing to fluxes
through a given cell-face should be centered at the center of that cell-face.
Moreover, a monotinization of derivatives is essential for the numerical
stability of the overall algorithm as soon as cosmic rays are coupled to the
gas dynamics through the $\bm{\nabla} p_{\rm cr}$ term in the equation of gas
motion. We apply the following formulae for numerical derivatives needed for
the the flux components through the 1-faces:

\begin{equation}
(\nabla e_{\rm cr})_{(i,j,k)}^{\rm 1f} =
[(d_x e_{{\rm cr}})_{(i,j,k)},
(d_y e_{{\rm cr}})_{(i,j,k)},
(d_z e_{{\rm cr}})_{(i,j,k)}],
\end{equation}
where 
\begin{equation}
(d_x e_{{\rm cr}})_{(i,j,k)}=(e_{{\rm cr}(i,j,k)}-e_{{\rm cr}(i-1,j,k)})/
(x_{(i)}-x_{(i-1)} ),
\end{equation}
\begin{eqnarray}
(d_y e_{{\rm cr}})_{(i,j,k)}&=&0.25(d_{y,l} e_{{\rm cr}}+d_{y,r} e_{{\rm cr}}) 
\label{dyecr}\\
   & &\times (1 + \sign(1, d_{y,l} e_{{\rm cr}} \ d_{y,r} e_{{\rm cr}})),\nonumber
\end{eqnarray}
\begin{eqnarray}
(d_z e_{{\rm cr}})_{(i,j,k)}&=&0.25(d_{z,l} e_{{\rm cr}}+d_{z,r} e_{{\rm cr}})
\label{dzecr}\\
  & &\times (1 + \sign(1, d_{z,l} e_{{\rm cr}}\ d_{z,r} e_{{\rm cr}})),\nonumber
\end{eqnarray}
and the left and right derivatives used in the above formulae are given by
\begin{eqnarray}
d_{y,l} e_{\rm cr}  &=&0.5((e_{{\rm cr}(i-1,j  ,k)} 
                       + e_{{\rm cr}(i ,j  ,k )})  \\
                  & &   -(e_{{\rm cr}(i-1,j-1,k)}   
           + e_{{\rm cr}(i ,j-1,k )}))/(y_{(j)}-y_{(j-1)} ),\nonumber \\
d_{y,r} e_{\rm cr}  &=&0.5((e_{{\rm cr}(i-1,j+1,k)} 
                       + e_{{\rm cr}(i ,j+1,k )}) \\
                  & &   -(e_{{\rm cr}(i-1,j ,k) }
           + e_{{\rm cr}(i ,j  ,k )}))/(y_{(j+1)}-y_{(j)} ),\nonumber\\
d_{z,l} e_{\rm cr}  &=&0.5((e_{{\rm cr}(i-1,j ,k )} 
                       + e_{{\rm cr}(i ,j  ,k )}) \\
                  & &   -(e_{{\rm cr}(i-1,j,k-1)} 
                 + e_{{\rm cr}(i ,j ,k-1)}))/(z_{(k)}-z_{(k-1)} ),\nonumber\\
d_{z,r} e_{\rm cr}  &=&0.5((e_{{\rm cr}(i-1,j,k+1)} 
                       + e_{{\rm cr}(i ,j ,k+1)})  \\
                  & &   -(e_{{\rm cr}(i-1,j,k  )} 
           + e_{{\rm cr}(i ,j,k  )}))/(z_{(k+1)}-z_{(k)} ).\nonumber
\end{eqnarray}
The monotonized derivatives used in formulae (\ref{dyecr}) and (\ref{dzecr})
reduce to centered derivatives if signs of left and right derivatives  are the
same. The cosmic ray energy fluxes through the cell-faces in the $y$ and $z$-
directions are constructed in the analogous way. The cosmic ray diffusive
fluxes through each cell-face can be now computed according to the formula
(\ref{cr-flux}). 

The standard stability analysis (see eg. Fletcher 1991) imposes  the following
necessary stability condition for the explicit numerical
solutions of the diffusion equation:
\begin{equation}
\Delta t \leq C_{{\rm cr}} \frac{(\min(\Delta x,\Delta y,\Delta z)^2}{K},
\label{timestep}
\end{equation}
where $C_{{\rm cr}}=0.5$ is the Courant number corresponding to the diffusion
problem. Numerical tests show that a
slightly lower value, namely 0.3 suits better for the current problem.

The above limitation for the timestep,  which is quadratic in the cell size,
together with the application of  monotonized derivatives implemented according
formulae (\ref{dyecr}) and (\ref{dzecr}), ensures a stable numerical scheme for
the active cosmic ray transport in the ZEUS code.

The standard types of boundary conditions  for the cosmic ray component can be
implemented in the similar way as for the internal energy, except the outflow
boundary condition. In case of the internal energy the outflow boundary
condition relies on the replication of the contents of the starting and ending
cells to the adjacent ghost zones. This kind of the outflow boundary condition
is, however, not appropriate for the diffusing cosmic ray component, since
replication of cell contents means no gradients across the boundary. If
gradients of cosmic ray energy vanish then only advection and no diffusion is
possible across the boundary. In order to make it possible for cosmic rays to
diffuse across the boundary of the physical domain we first compute gradients
of cosmic ray energy density across the boundary and then perform a linear
extrapolation of the cosmic ray energy density from the interior cells to the
ghost zones. 

\section{Test problems}

In this section we present several tests of our new numerical algorithm for the
cosmic-ray diffusion-advection problem. 

In the following considerations we apply units which are convenient for the
investigations of the dynamics of ISM on large spatial scales. The unit of
length and time are 1 pc and 1 Myr, respectively. The unit of velocity is $1
\pc \Myr^{-1} \simeq 1 \km \s^{-1}$. The density is given in $\cm^{-3}$
corresponding to the number density of hydrogen atoms. The unit of the magnetic
field is  $1\mu{\rm G}$. The diffusion coefficient is expressed in units of $1
\pc^2 \Myr^{-1}$. The realistic parallel  diffusion coefficients for the cosmic
ray transport as estimated by Strong and Moskalenko (1998), is based on the
GALPROP model: $K_\parallel = 6 \times 10^{28} \cm^2 \s^{-1} = 2 \times 10^5
\pc^2 \Myr^{-1}$. In our numerical tests we apply diffusion coefficients
ranging from $10^2 \div 10^5 \pc^2 \Myr^{-1}$. 

We present test problems which are performed in the
computational box of physical size $1000 \pc \times 1000 \pc \times 500 \pc$
with a spatial resolution of $100 \times 100 \times 50$ grid zones in the $x$, $y$ and
$z$ directions, respectively. Periodic boundary conditions are applied on
all the domain boundaries.

\subsection{Passive cosmic ray transport in one dimension}

\begin{figure}
\centerline{\includegraphics[width=\columnwidth]{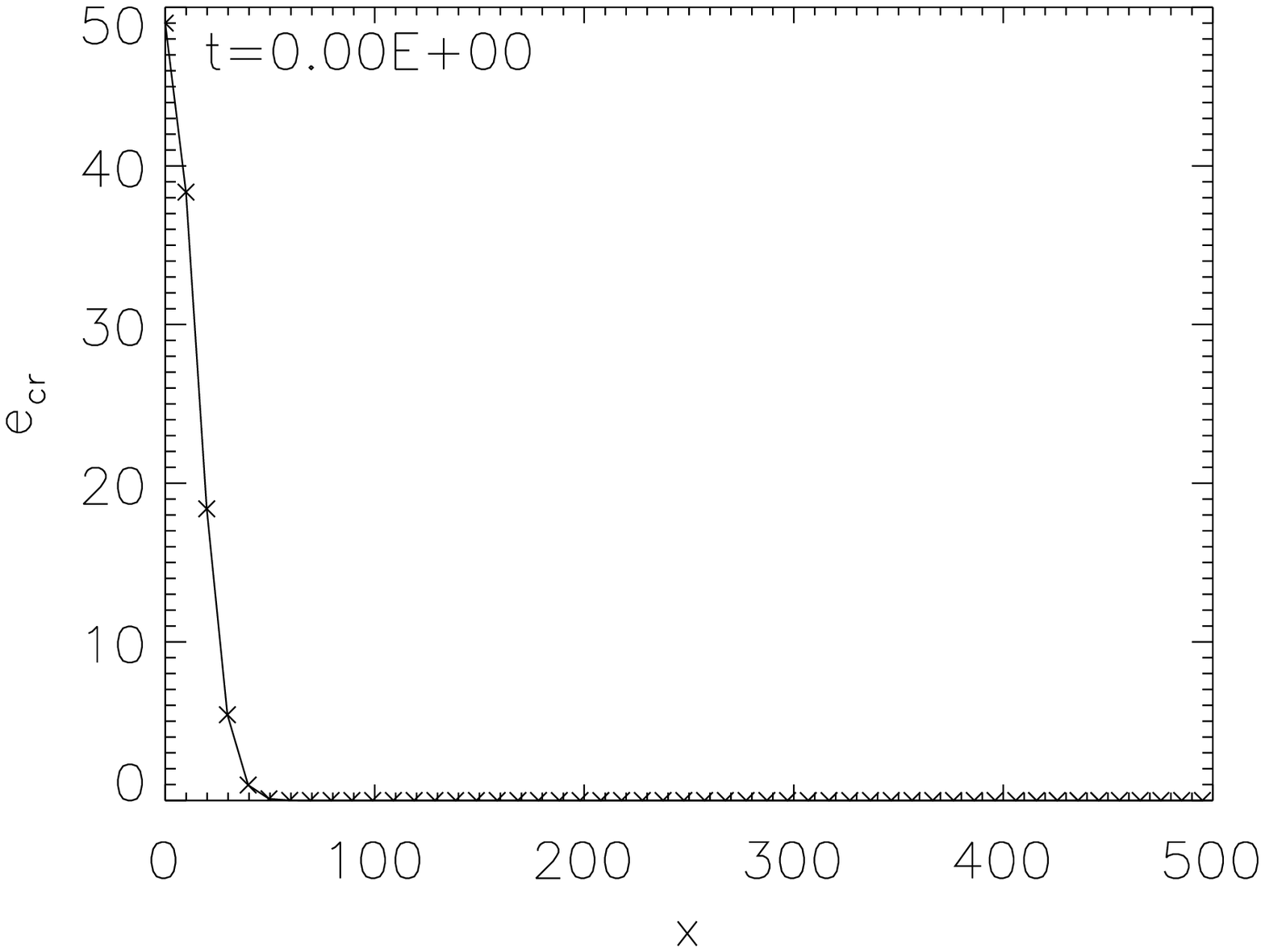}}
\centerline{\includegraphics[width=\columnwidth]{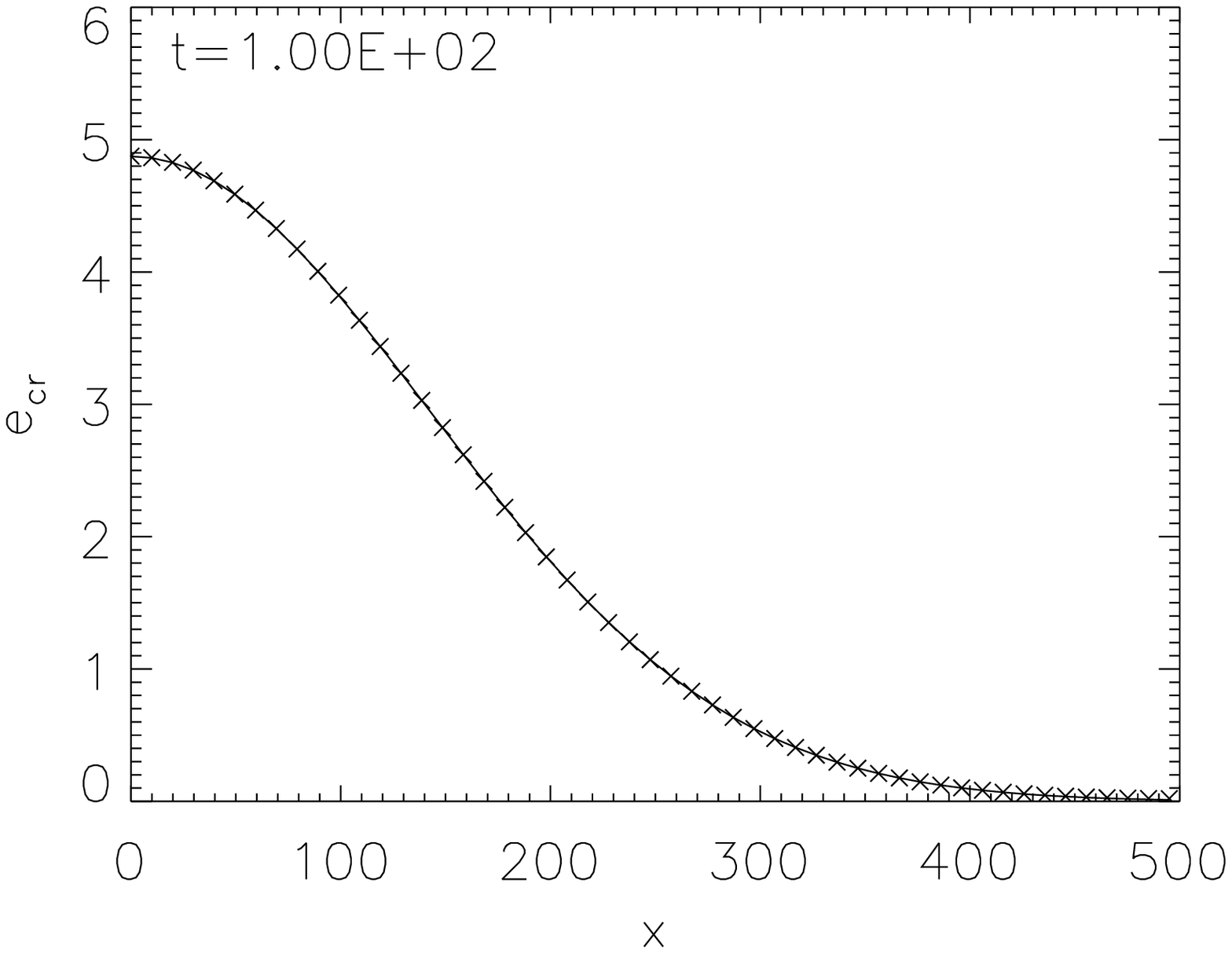}} 
\caption{Comparison of numerical solutions of the 1D, pure diffusion problem to
analytical solutions for $t= 0$ (upper panel) and $t=100$ (lower panel). 
Continuous lines denote analytical
solutions for different times and crosses represent numerical solutions.
}
\label{fig1}
\end{figure}

As a first step we perform a test of passive cosmic ray transport along the 
magnetic field, which is directed along the $x$-axis, assuming $K_\perp=0$ and
a static medium, i.e $\bm{v}=0$. The passive transport means  that 
the cosmic ray gas has no dynamical influence onto the motion of thermal gas. This
can be achieved by neglecting the  $\bm{\nabla} p_{\rm cr}$ in  the equation of gas
motion. In case of a static distribution of gas and a uniform magnetic field
parallel to the $x$-axis the cosmic ray transport equation reduces to a
one-dimensional diffusion problem, described by the equation

\begin{equation} 
\pder{e_{\rm cr}}{t}{} = K \pder{e_{\rm cr}}{x}{2}.
\label{diffeq} 
\end{equation} 
Assuming that the initial condition is given by
the cosmic-ray distribution 
\begin{equation} e_{\rm cr0} = A
\exp{\left(-\frac{x^2}{r_0^2}\right)}, 
\end{equation} 
where $r_0$ denotes the initial
half-width of the Gaussian profile, we expect that the numerical solution
should be close to the following analytical solution at any time
\begin{equation} 
e_{\rm cr}(x, t) = A \sqrt{\frac{r_0^2}{r_0^2 + 4 K t}}  
  \exp{\left(-\frac{x^2}{r_0^2 + 4 K t} \right)} 
  \label{gauss1d} 
\end{equation} 
The comparison of the numerical solutions
with the analytical solution (eq. 20) is shown in Fig.~\ref{fig1}. A perfect
consistency of the both analytical and numerical solutions is obvious.

\subsection{Passive cosmic ray transport along an inclined magnetic field}

The next test for the diffusive cosmic-ray propagation is to check if the
simulated diffusion follows the analytical solution in case of an inclined magnetic field.
We set up the values of $K_\parallel = 100$ and
$K_\perp = 0$ and perform simulations for $B_x = B_y \neq 0$ and $B_z=0$. 

\begin{figure}
\centerline{\includegraphics[width=\columnwidth]{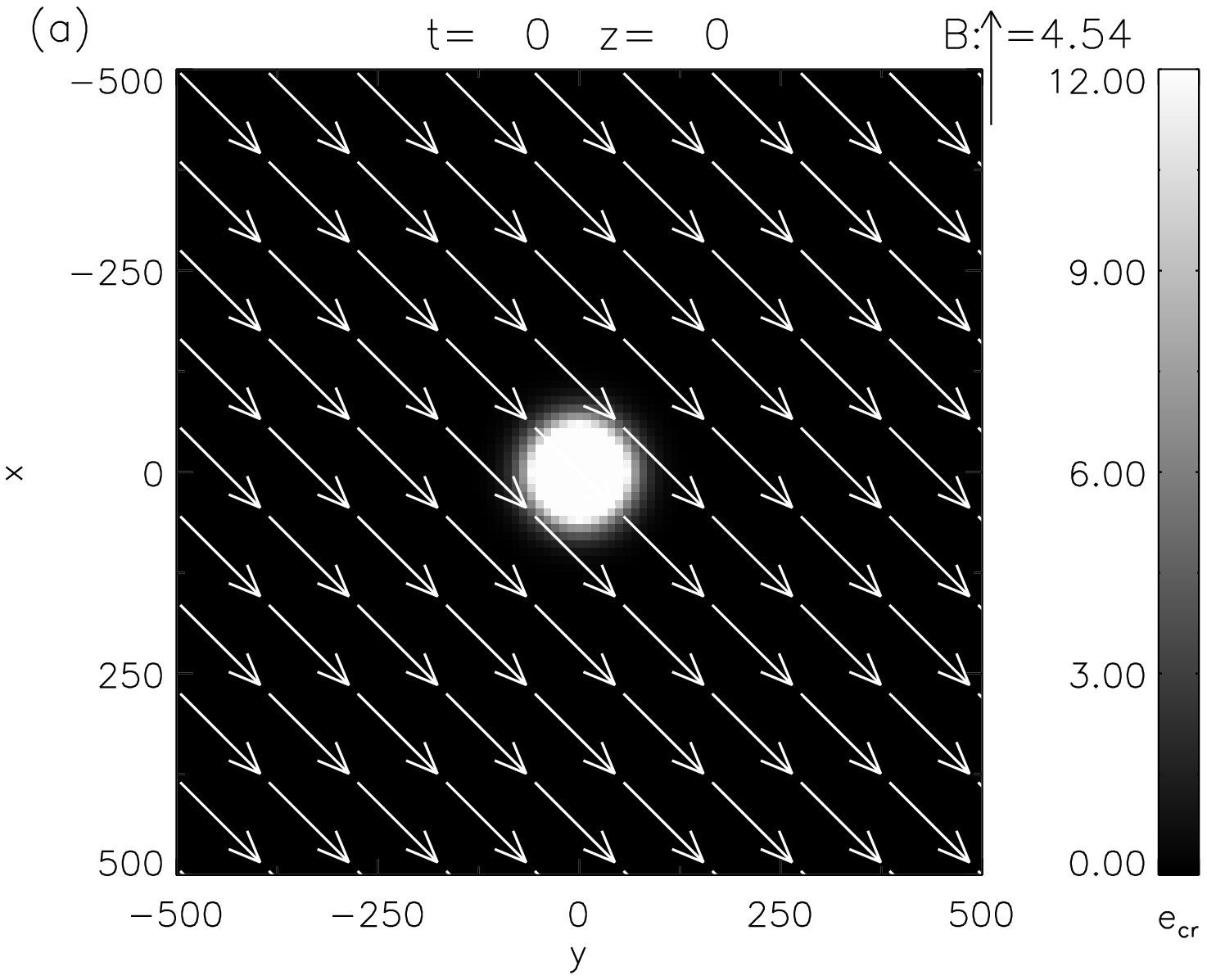}}
\centerline{\includegraphics[width=\columnwidth]{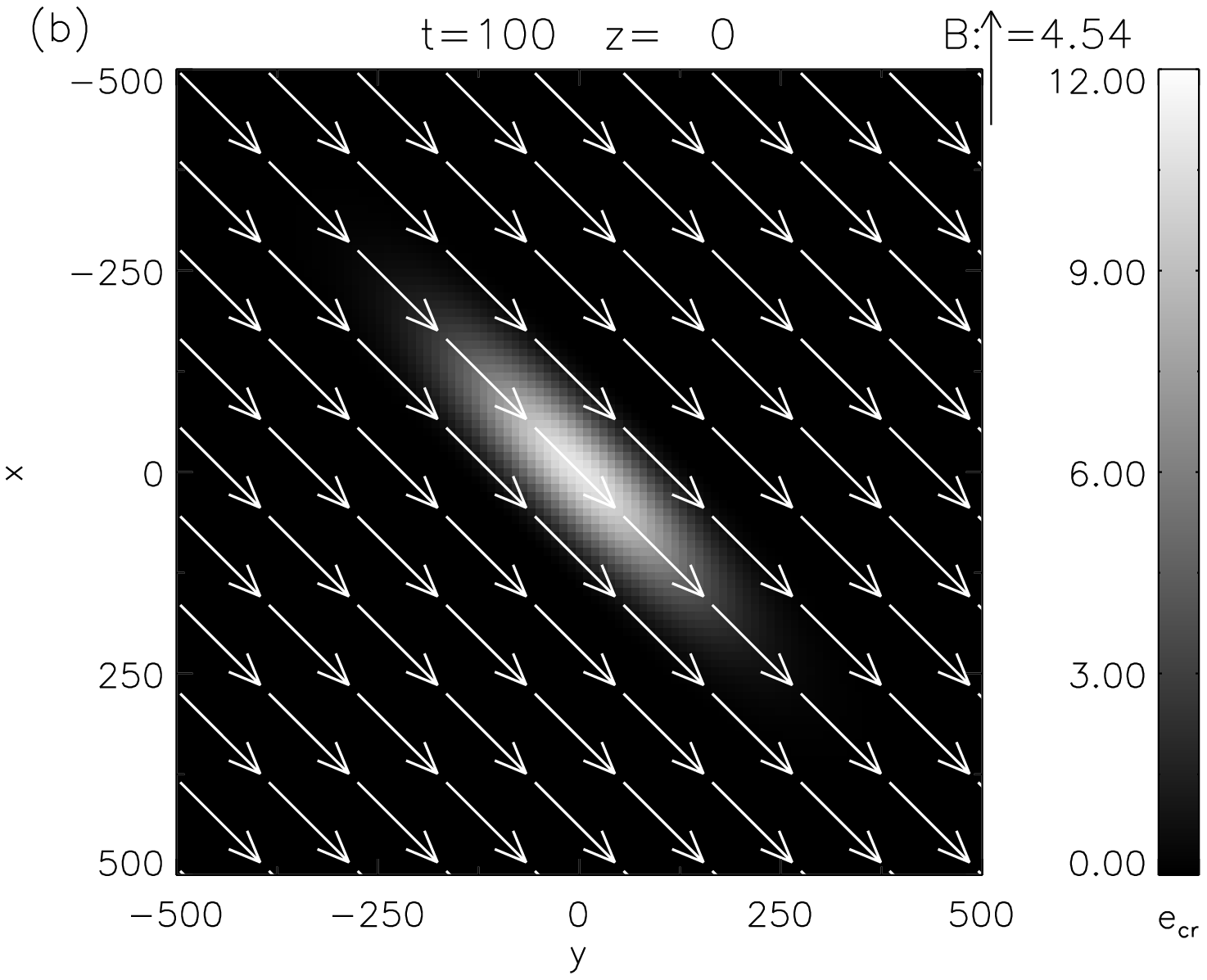}}
\caption{Diffusion of cosmic rays along an inclined magnetic field: (a) the initial
spheroidal distribution of $e_{\rm cr}$ at $t= 0$ and (b) the ellipsoidal
distribution at $t=100$. 
}
\label{fig2}
\end{figure}

Fig.~\ref{fig2} presents two snapshots of cosmic ray distribution  at $t=0$
and $t=100$. It is apparent that qualitatively the anisotropic transport of
cosmic ray energy proceeds along the magnetic field according to our expectations.

\begin{figure}
\centerline{\includegraphics[width=\columnwidth]{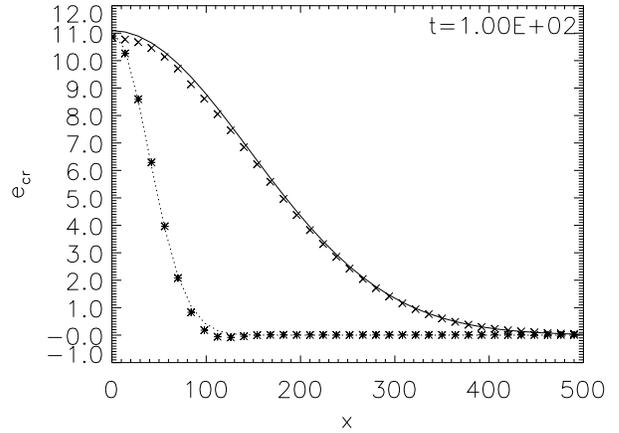}}
\caption{
Diffusion of cosmic rays along an inclined magnetic field at  $t=100$.   Two cuts
of the ellipsoid shown in Fig.~\ref{fig2} are made (b) along the major axis 
(crosses) and along the minor axis (asterisks). The full and dotted lines
represent the corresponding cuts of the fitted 2D Gaussian profile.
}
\label{fig3}
\end{figure}

However, a more precise evaluation of the numerical algorithm can be performed by
fitting a 2D-Gaussian function for the distribution of cosmic ray energy density
in the $xy$-plane. The diffusion along the $z$-direction is absent in the current
setup. The fitting procedure is performed with the aid of IDL routine {\em
gauss2dfit}, which returns parameters of a 2D-Gaussian distribution, i.e the
semi-axis of the ellipsoid, as well as the amplitude of the peak and the level
of the constant background distribution of cosmic ray energy.  The results of
the fitting are shown in  Fig.~\ref{fig3}, where  two 1-d slices along the
major and minor axis of the ellipsoid (crosses and asterisks respectively) are
shown together with the corresponding curves calculated on the base of the derived
parameters of the ellipsoid. 

In the present case the fitted values of the parallel and perpendicular widths
of the 2D Gaussian profile and the fitted amplitude are  $r_{\parallel f}=
206.429$, $r_{\perp f}= 54.705$,  $A_f = 11.096$. Since $K_\perp=0$, the
corresponding parameters of the exact solution can be derived from formula
(\ref{gauss1d}) representing the time evolution of 1D Gaussian profile. The
exact analytical solution gives $r_{\parallel a}= \sqrt{r_0^2 + 4 K t}=206.155$
and $A_a = A/\sqrt{r_0^2 + 4 K t} = 11.884$. If no perpendicular diffusion is
present, the perpendicular widths should not change, i.e. $r_{\perp
a}= 50.000$. 

We note that the parallel width of the 2D Gaussian profile of the simulated cosmic
ray distribution coincides well with the analytical solution. 
However, there are two noticeable effects of the limited accuracy of our numerical
algorithm. The first effect is the broadening of the perpendicular widths
$r_\perp$ by about 10\% of the original value.  The broadening leads to a
lowering of the peak amplitude $A_f$ with respect to the exact solution since
the total cosmic ray energy is conserved in absence of gas flows.  The second
effect of the limited numerical accuracy affects the formation of regions with
negative values of cosmic ray energy density at the base of steep sides of the
ellipsoid.

The depth of the $\ecr$ deficit (currently  equal to $-0.09123$ at $t=100$) is
strongly dependent on the grid resolution and the steepness of the initial
cosmic ray energy distribution across the magnetic field lines. The presence of
negative CR energy density regions may lead to significant numerical artifacts
as soon as cosmic rays are coupled to the equation of gas motion. Therefore the
spatial resolution of the grid in conjunction with the magnitude of the cosmic ray
gradients is an issue of primary importance. The deficit vanishes in 
proportion to the grid resolution, however there is a limited freedom of
reducing the size of the grid cell due to a drastic reduction of the timestep
given by formula  (\ref{timestep}). 

An alternative way is to incorporate an
explicit perpendicular diffusion given by $K_\perp$ equal to a few up to a few
10 percent of $K_\parallel$. This procedure can be physically justified by
studies of cosmic ray transport in turbulent magnetic fields (Giaccalone and
Jokipii 1999). Another way to eliminate the spots of negative cosmic ray
energy density is to add a smooth background of cosmic rays prior to the
injection of very localized portions of cosmic ray energy. This is also a
physically justified procedure, since e.g. the smooth background of cosmic rays
is present in the ISM.

\subsection{Active cosmic ray transport}

After testing the passive diffusion of cosmic rays we can now describe the
active propagation of cosmic rays within the overall setup similar to that of
the previous subsection. In the present configuration we apply the uniform
background of cosmic rays resulting from the assumption that cosmic ray
pressure is equal to the magnetic and gas pressure \begin{equation} p_{\rm tot}
= p_{\rm gas} + p_{\rm mag} + p_{\rm cr}  \label{ptot} \end{equation} where 
\begin{eqnarray} p_{\rm mag} = \alpha_p p_{\rm gas}\\ p_{\rm cr} = \beta_p
p_{\rm gas} \end{eqnarray} and $\alpha_{\rm p}$ and $\beta_{\rm p}$ are in
general constants of  the order of 1 in the ISM. In the present case we adopt
$\alpha_{\rm p}=\beta_{\rm p}=1$. The constant scaling factors between
cosmic ray and gas background pressures as well as between magnetic and gas
pressures is adopted for testing purposes only. This approach is useful
especially in linear studies of  Parker instability and follows the original
work by Parker (1966, 1967).  In general, one can start with any arbitrarily
chosen initial magnetic field and cosmic ray distributions. The corresponding
background cosmic ray energy density in our units is $e_{\rm cr} = (\gamma_{\rm
cr} - 1)^{-1} \alpha_p c_{\rm si}^2 \rho_0 = 147 $, where $c_{\rm si} = 7 \km
\s^{-1}$ denotes the isothermal sound speed and  $\rho_0 = 1$ is the background
gas density. 

In order to investigate the coupling of cosmic rays to gas and magnetic field
we switch on the $\bm{\nabla} p_{\rm cr}$ term in the equation of gas motion 
and inject about 50 \% 
of  $10^{51}\erg$ 
of the kinetic output of the SN
explosion. In our scaled units this corresponds to the initial peak amplitude
of the Gaussian CR distribution equal to 100 times the background thermal 
energy density, i.e $100 c_{\rm si}^2 \rho_0 = 4900$. Currently we apply a
rather small diffusion coefficient of cosmic rays $K_{\parallel} = 100$
(corresponding to $3 \cdot 10^{25} \cm^2\s^{-1}$) and  $K_{\perp} = 4$  in
order to illustrate the effects of cosmic ray propagation qualitatively. 

\begin{figure}
\centerline{\includegraphics[width=\columnwidth]{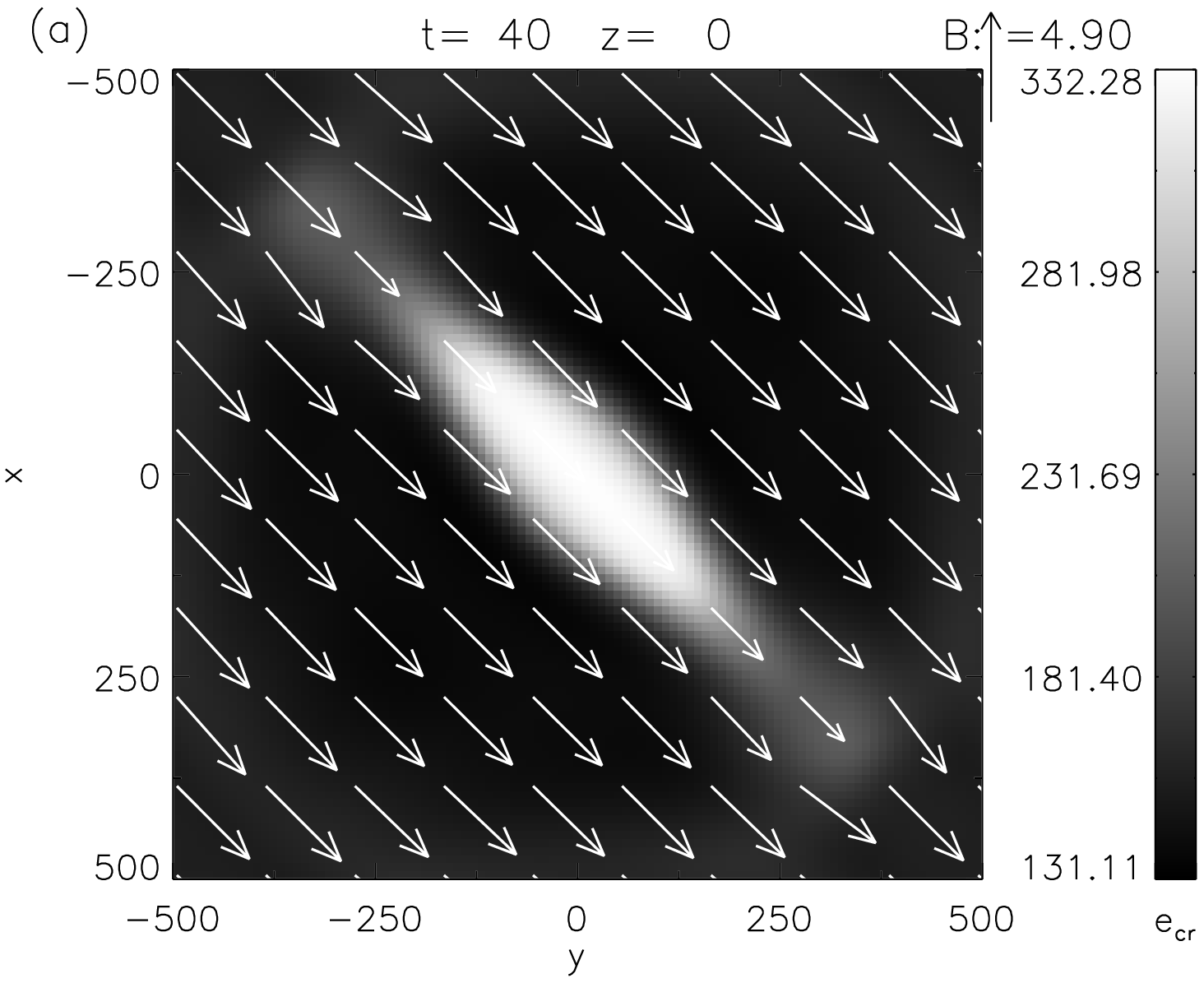}}
\centerline{\includegraphics[width=\columnwidth]{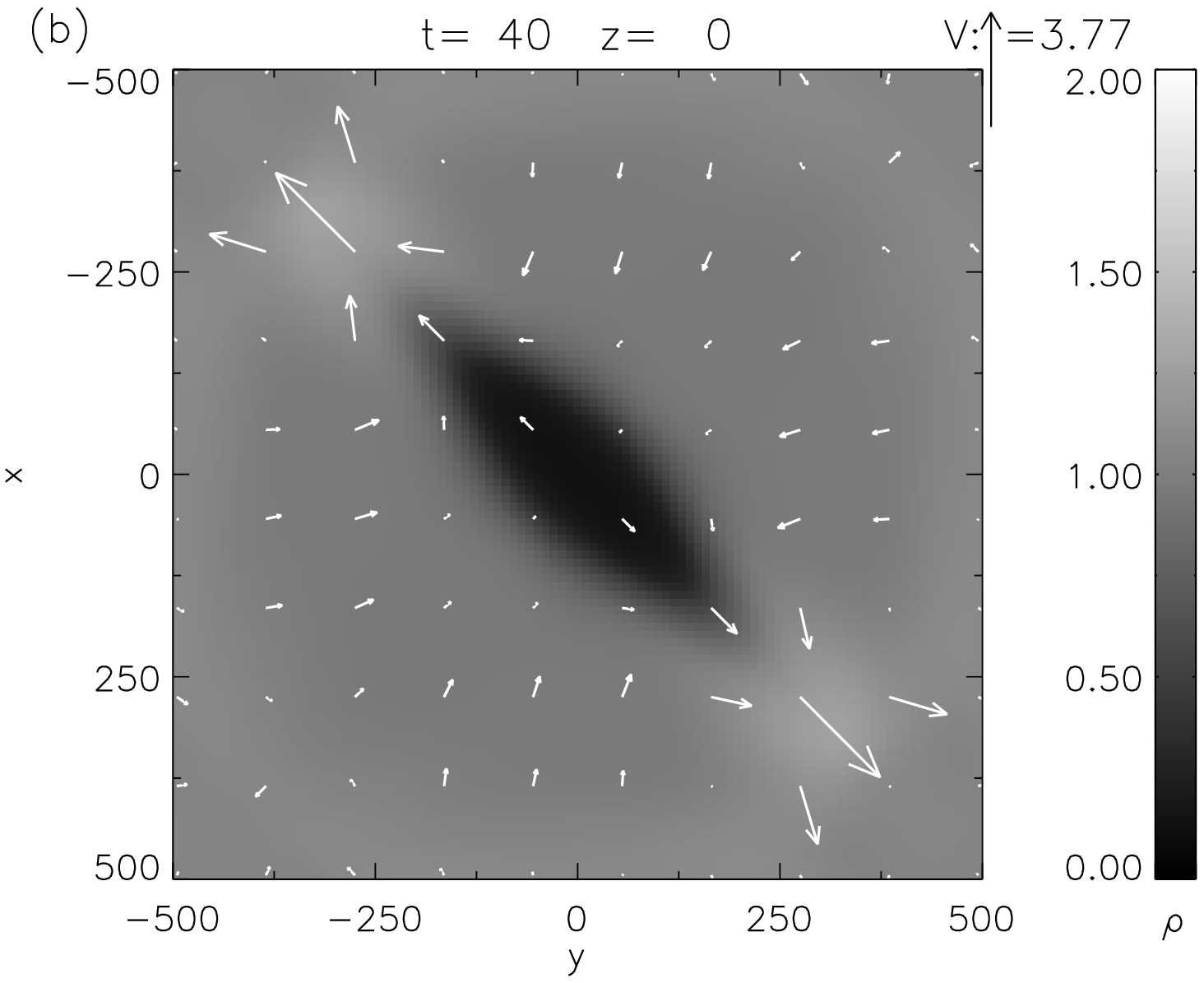}}
\caption{Diffusion of cosmic rays along an inclined magnetic field at $t=40$: (a) 
the cosmic ray energy density $e_{\rm cr}$ 
and magnetic field 
and (b) distribution of gas density and gas velocity.}
\label{fig4}
\end{figure}

In Fig.~\ref{fig4} we present the cosmic ray - magnetic field distribution (panel
a) and the density - velocity distribution (panel b) at $t=40$. The difference
between the passive and active cosmic ray transport is remarkable. First of all
the gradient of the cosmic ray pressure leads to the acceleration of gas. Due to
the effect of magnetic tension gas accelerates preferentially along the magnetic
field up to a few $\km \s^{-1}$. The outflow of gas from the cosmic
ray injection region together with the expansion forced by the cosmic ray pressure
implies the formation of a cavity in the gas distribution. On the other hand, gas
accumulates outside the cavity as it is visible in \ref{fig4} (b), in the
upper-left and lower-right corners of the graphic.

The gas motion along the magnetic field lines  leads to an advection of cosmic rays
with the gas flows, which is noticeable in Fig.~\ref{fig4} (a) as an enhanced
cosmic ray energy density coinciding with the enhanced gas density in
Fig.~\ref{fig4} (b). The coupling of cosmic rays to the gas component implies
that the cosmic rays spread faster with respect to the passive (pure diffusion)
transport.   A broadening of the cosmic ray profile across the magnetic field is due
to the pressure of the cosmic ray gas and partially due to the imposed perpendicular
diffusion.

\section{Active cosmic ray transport in a vertically stratified atmosphere}

For the simulations of the cosmic ray transport  in a stratified atmosphere we
adopt a  physical domain and grid sizes $500 \times 1000 \times 1000$  pc
and $50 \times 100 \times 100$ grid zones, in $x$, $y$ and $z$ directions
respectively. We apply periodic boundary conditions to all the vertical domain
boundaries,  a reflection boundary condition to the lower domain boundary and
outflow condition to the upper boundary.

\begin{figure}
\centerline{\includegraphics[width=\columnwidth]{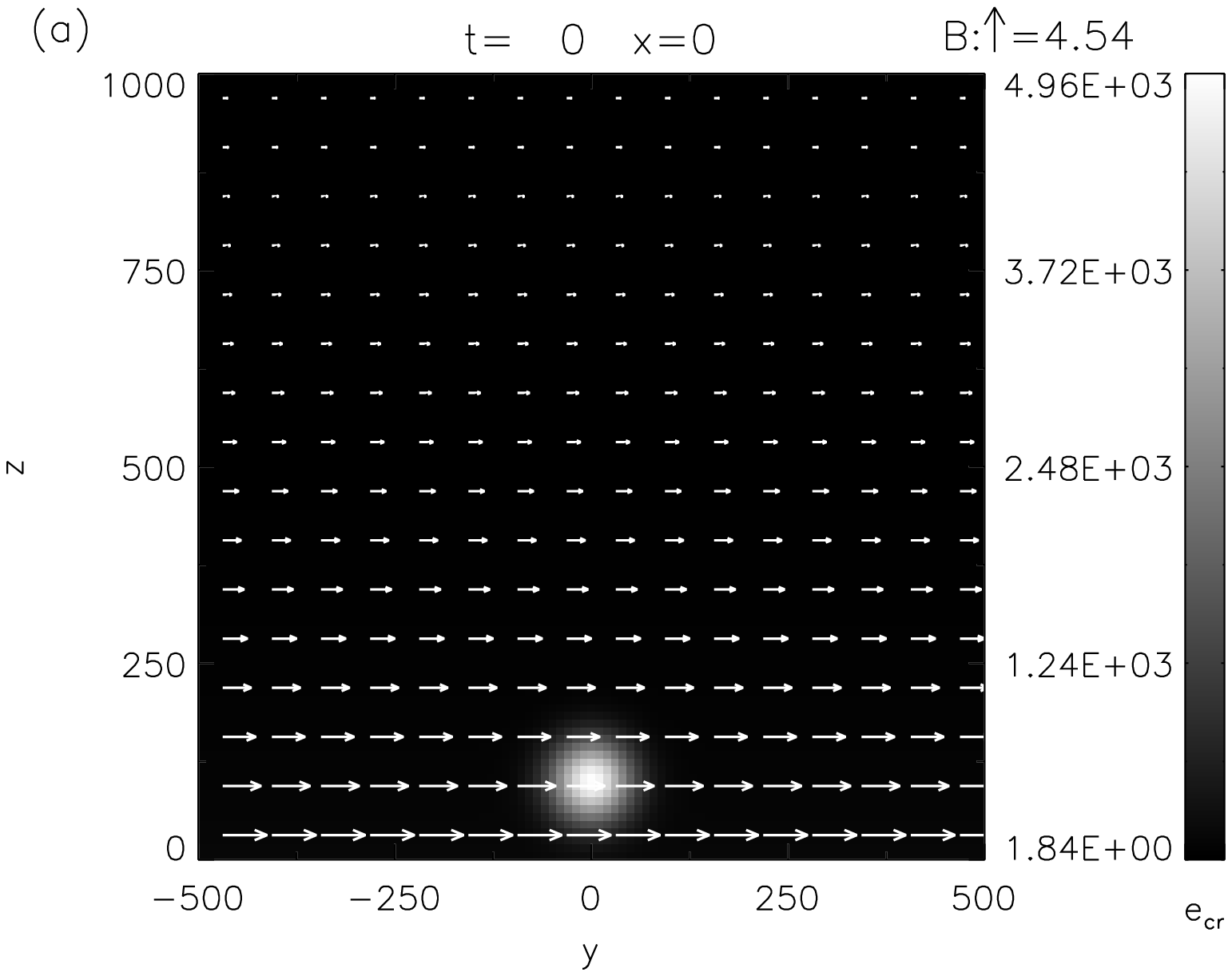}}
\centerline{\includegraphics[width=\columnwidth]{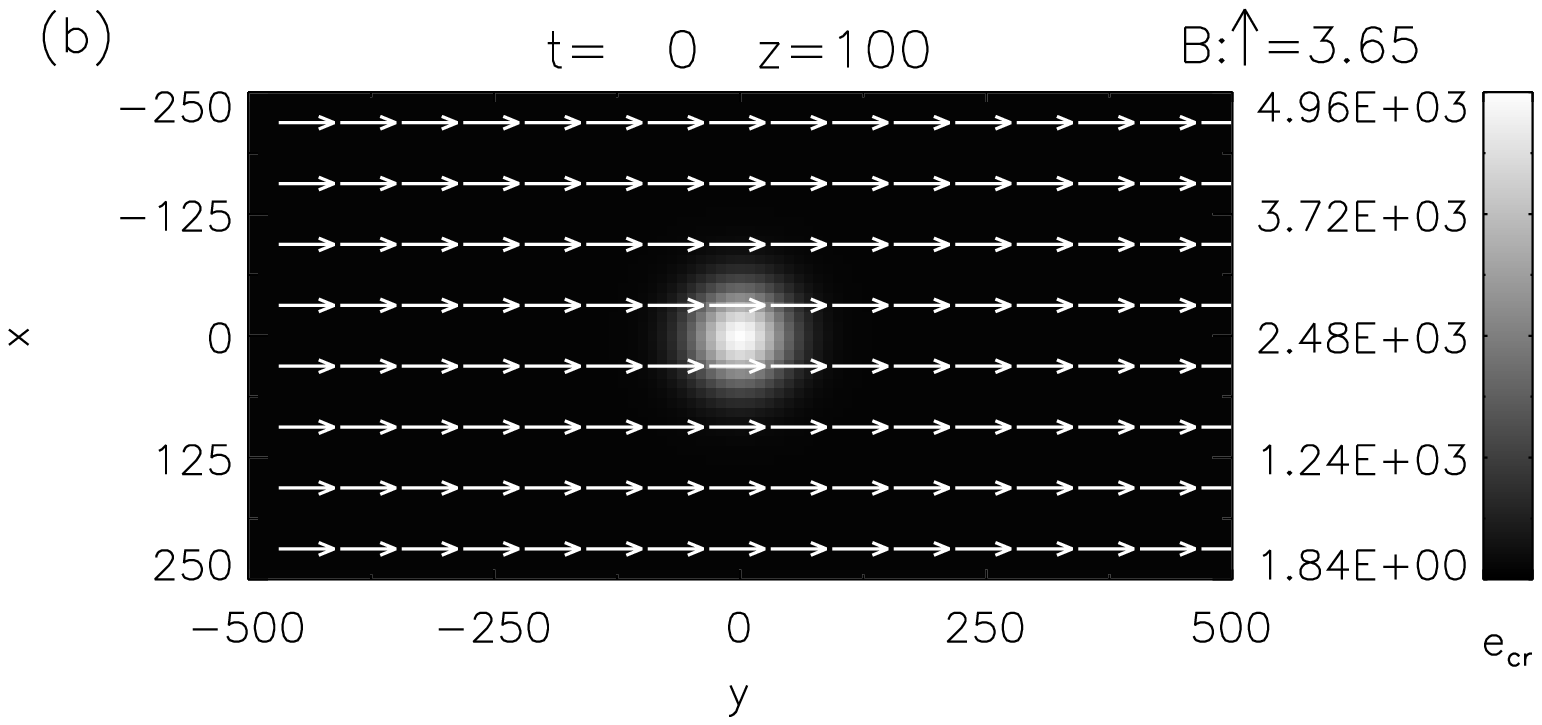}}
\caption{Geometry of the initial state stratified by vertical gravity: (a) the
slice in $yz$-plane showing the stratification of the background distribution of
cosmic rays and magnetic field along with the localized cosmic ray injection
region, (b) the initial, horizontal  magnetic field  is parallel to the 
horizontal and parallel to the $y$-axis.   } 
\label{fig5} 
\end{figure}

The goal of the present work is to incorporate the cosmic ray transport into
studies of the dynamics of a gravitationally  stratified interstellar medium.
In this section we perform an experiment similar to the ones presented in the previous
sections. However, in the present case a uniform vertical gravity is taken into
account for the construction of an initial equilibrium state.
The equilibrium fulfills the magnetohydrostatic force balance equation

\begin{equation}
-\der{p_{\rm tot}}{z}{} + g_z = 0,
\end{equation}
where $p_{\rm tot} = p(1 + \alpha_p + \beta_p)$ denotes the total pressure and
$g_z = -0.65 \pc \Myr^{-2} $ is the vertical, uniform gravitational
acceleration.  The center of cosmic ray injection is placed at $x=0$, $y=0$ and
$z=100$.   Two slices illustrating the geometry of the initial state are shown
in Fig.~\ref{fig5}.

\begin{figure}
\centerline{\includegraphics[width=\columnwidth]{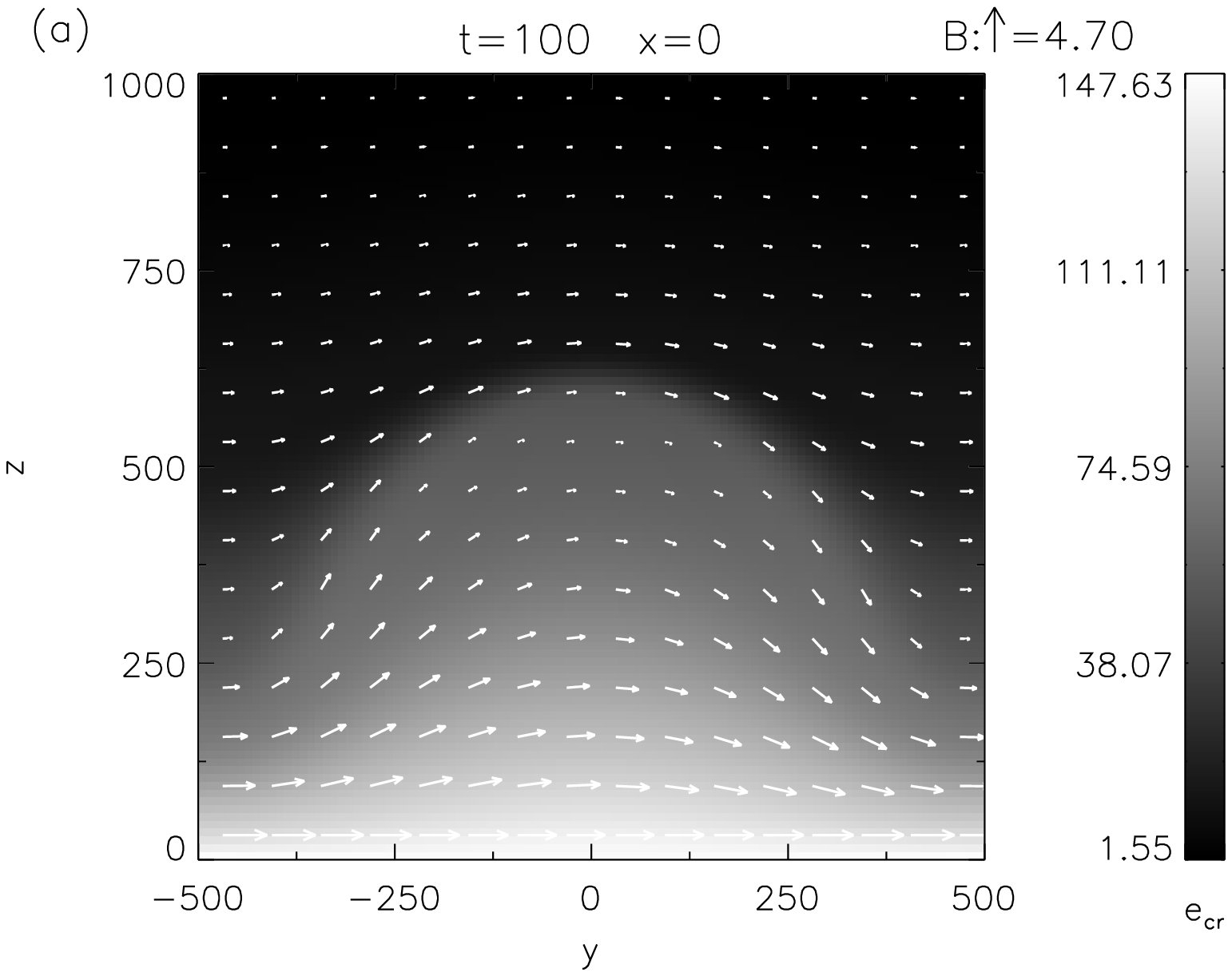}}
\centerline{\includegraphics[width=\columnwidth]{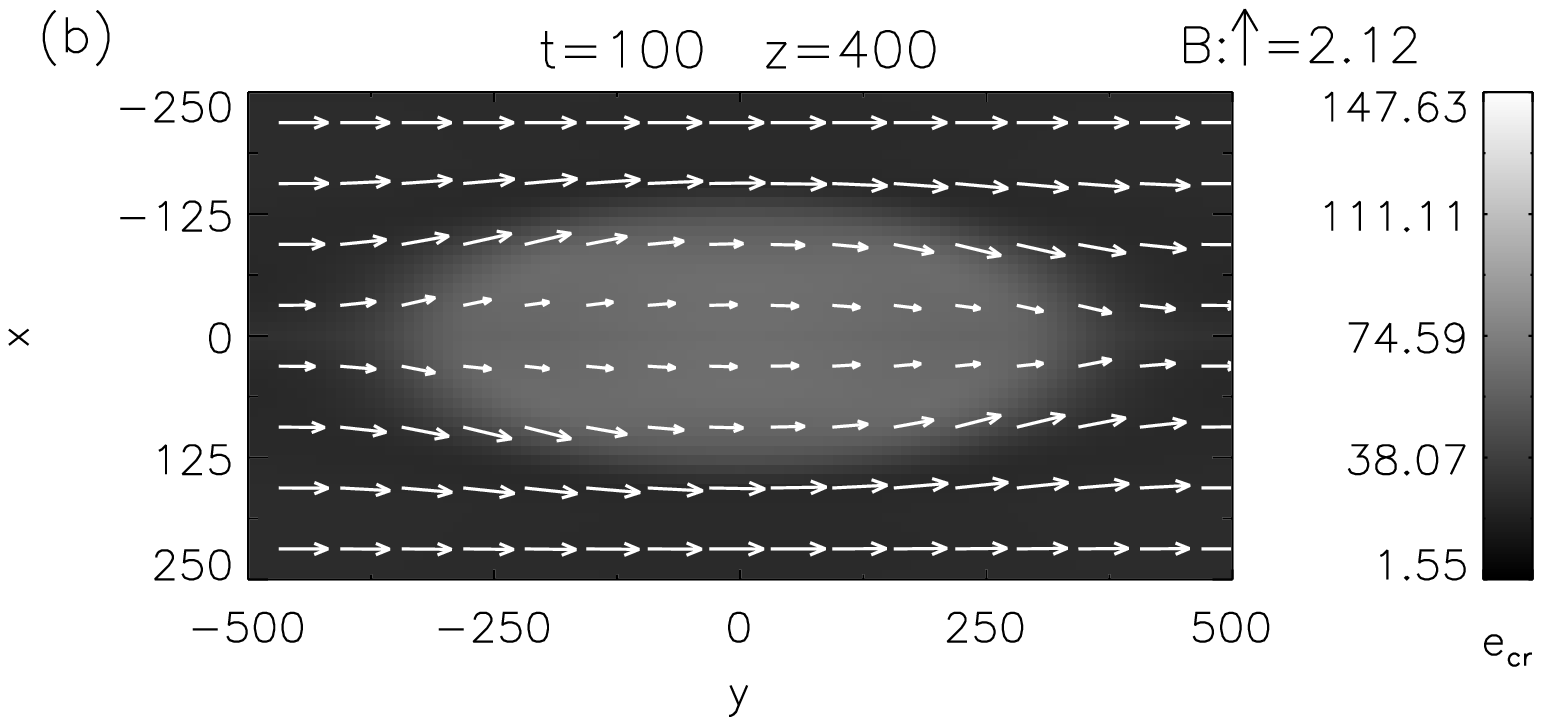}}
\centerline{\includegraphics[width=0.7\columnwidth]{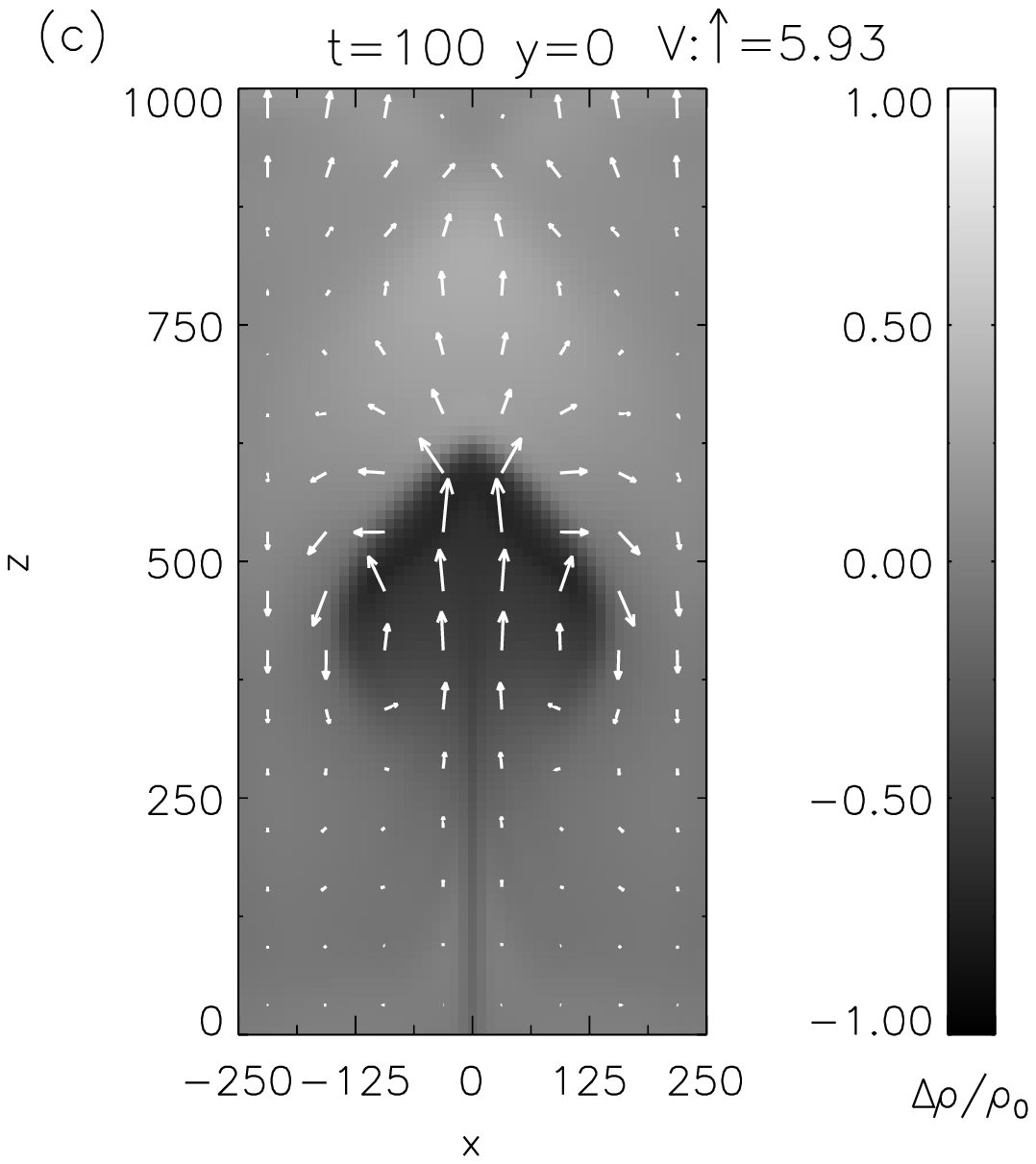}}
\caption{
Propagation of cosmic rays in a vertically stratified atmosphere for the
diffusion coefficient $K_\parallel=10^4$, after  the local injection of cosmic
ray energy: (a)  cosmic ray energy density and magnetic field in the $yz$-plane
at $x=0$ and (b) in the $xy$-plane at $z=400$,  (c) the density perturbation 
$\Delta \rho/ \rho_0$ and the gas velocity in the $xz$-plane.  } 
\label{fig6}
\end{figure}

\begin{figure}
\centerline{\includegraphics[width=\columnwidth]{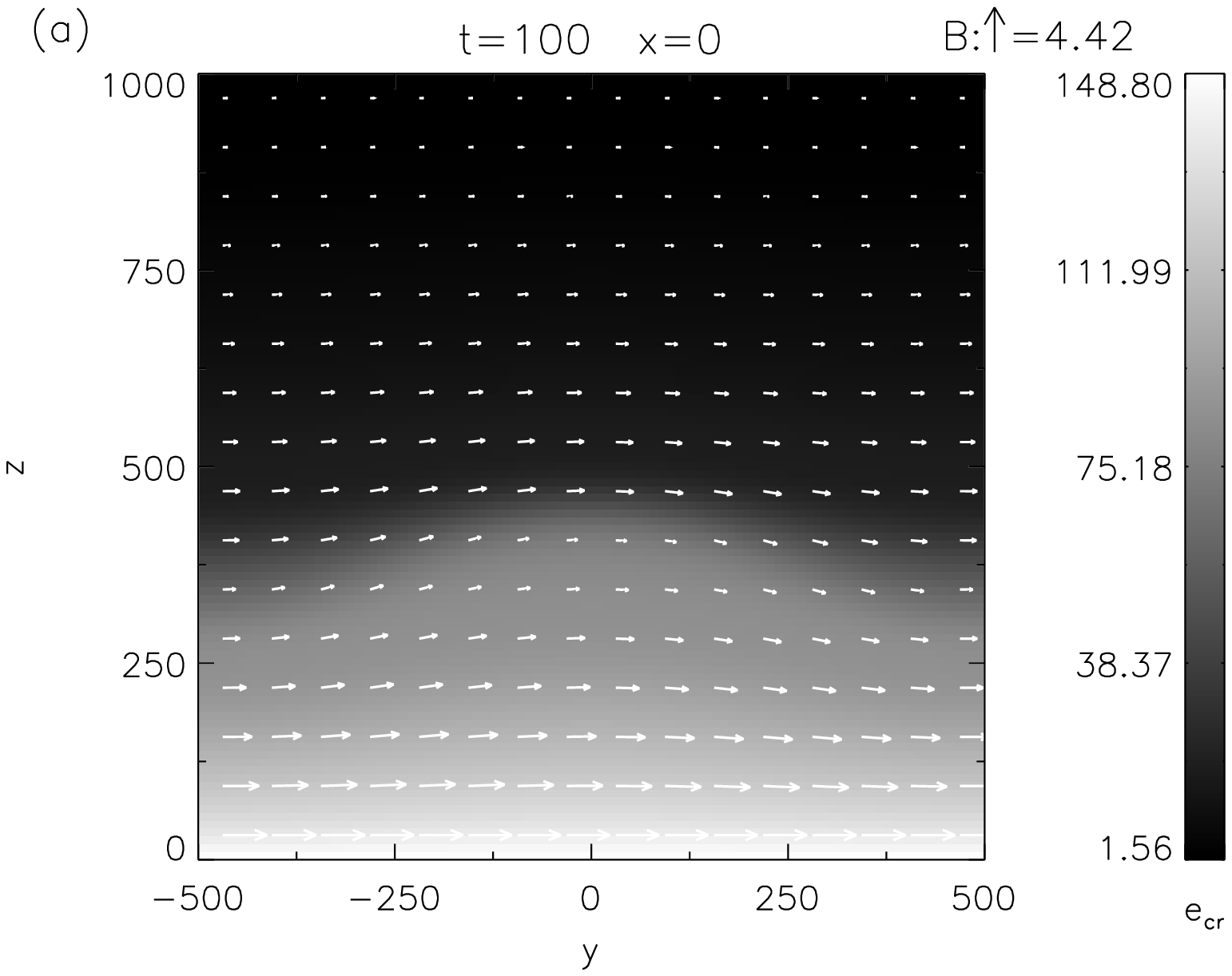}}
\centerline{\includegraphics[width=\columnwidth]{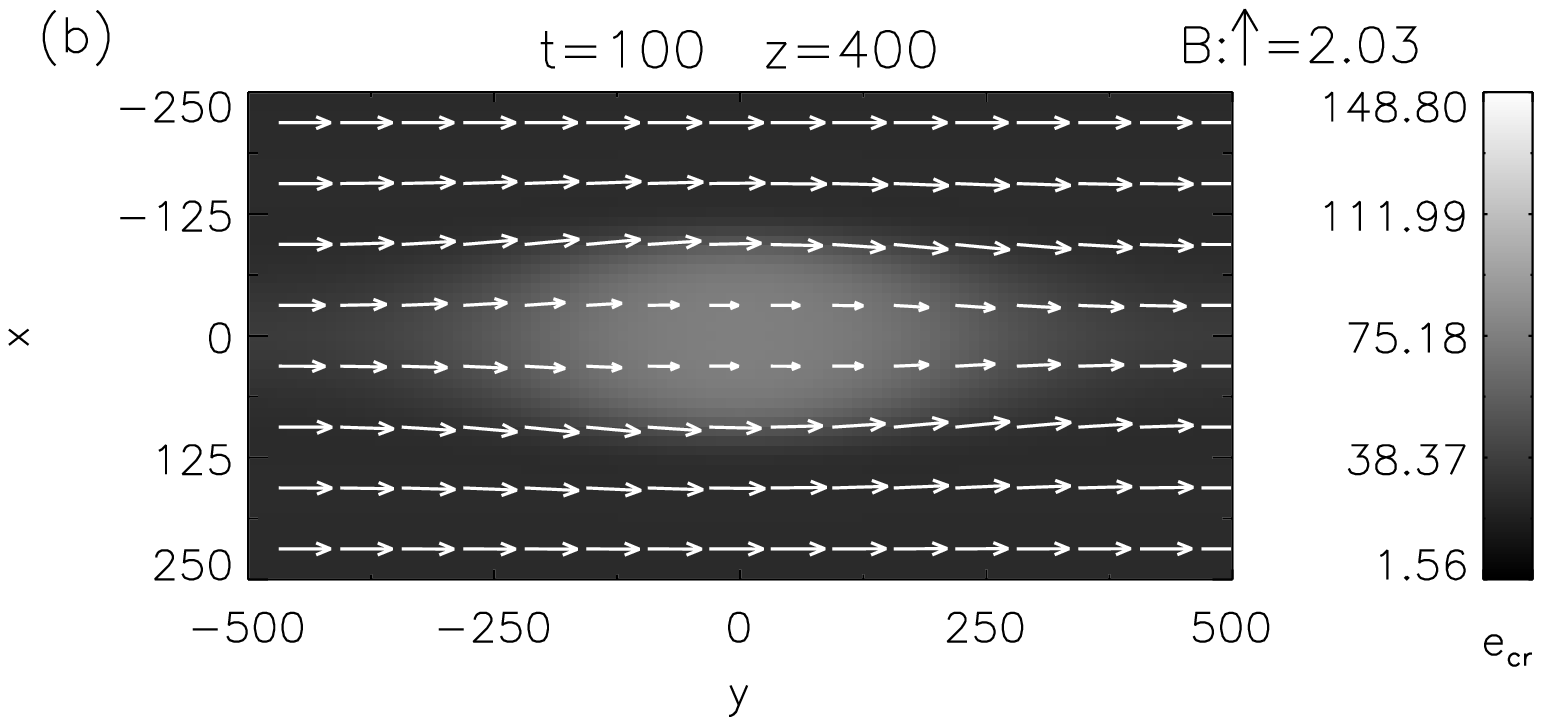}}
\centerline{\includegraphics[width=0.7\columnwidth]{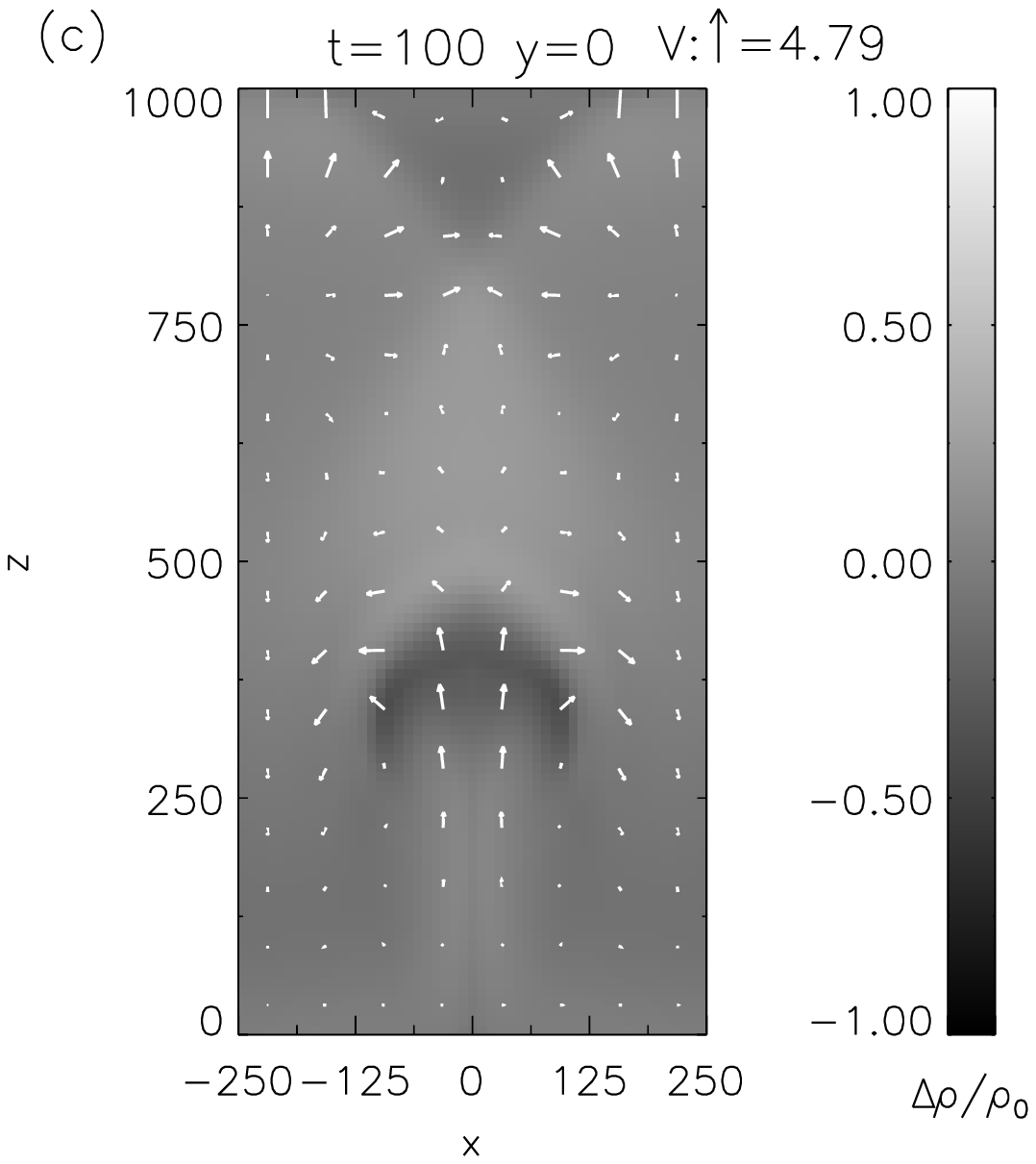}}
\caption{
Same as Fig.~\ref{fig6} for the
diffusion coefficient $K_\parallel=10^5$ at $t=100$.  
}
\label{fig7}
\end{figure}

\begin{figure}
\centerline{\includegraphics[width=\columnwidth]{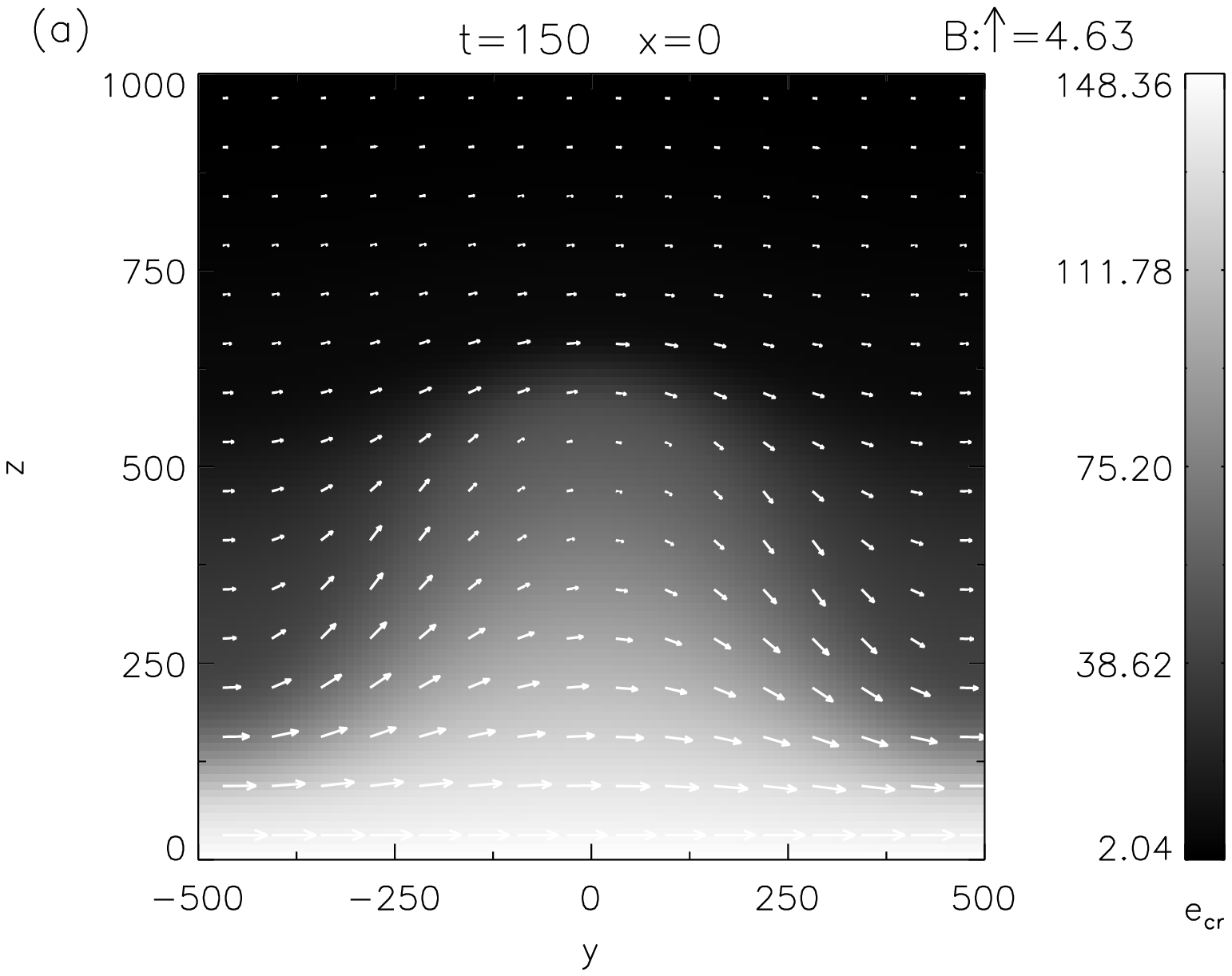}}
\centerline{\includegraphics[width=\columnwidth]{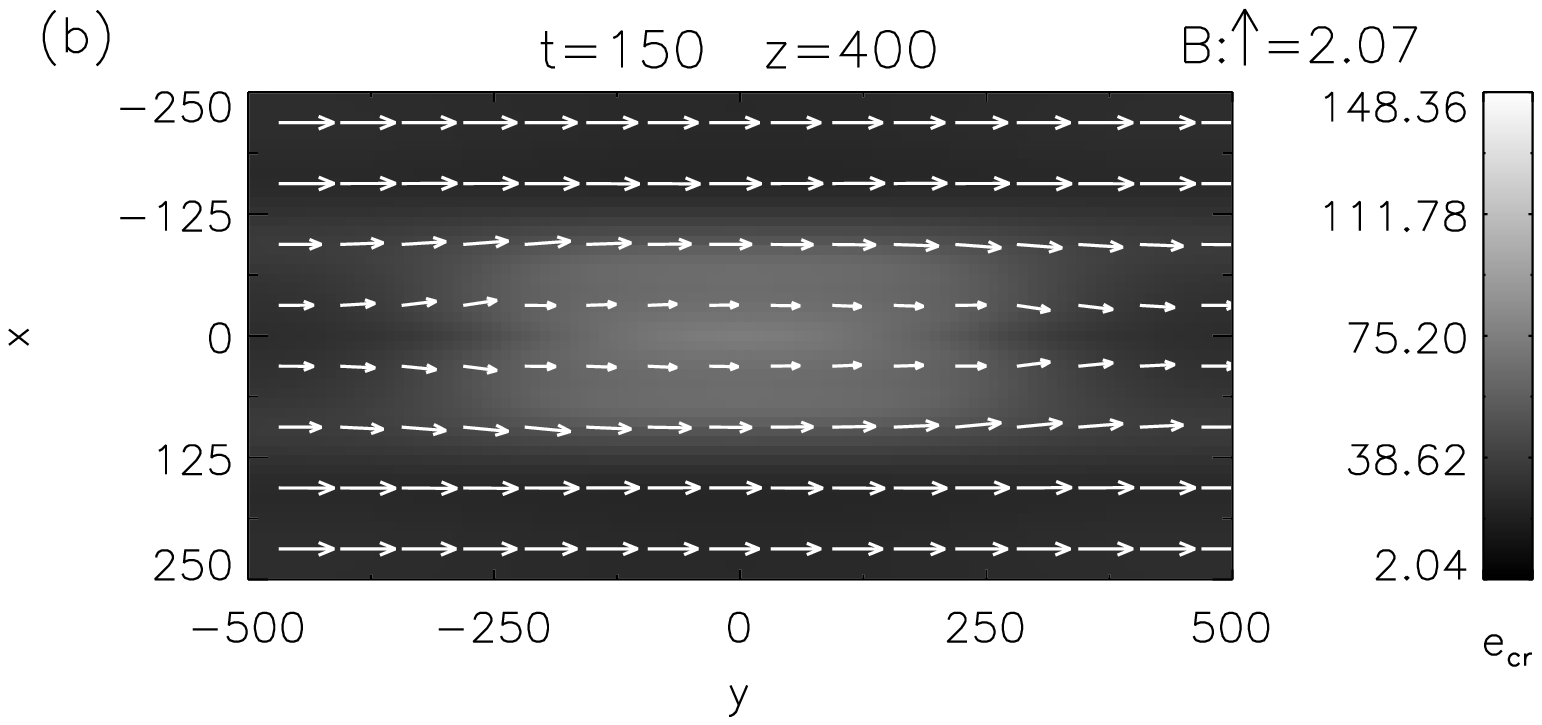}}
\centerline{\includegraphics[width=0.7\columnwidth]{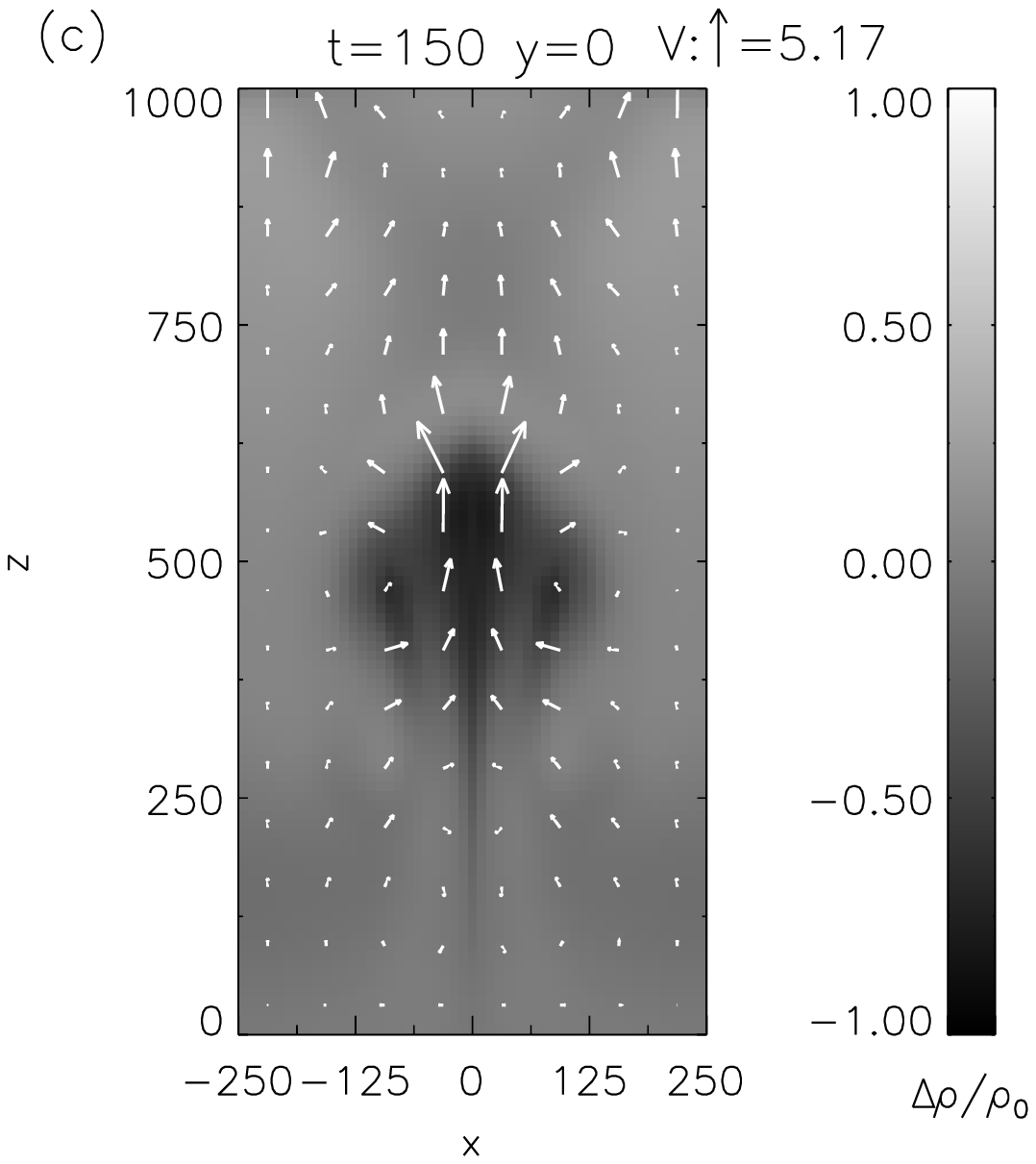}}
\caption{
Same as Fig.~\ref{fig7} at $t=150$.  
}
\label{fig8}
\end{figure}

Fig.~\ref{fig6} shows  the state of the system at $t=100$ in case of $K_\parallel
= 10^4$ (corresponding to $3\cdot 10^{27}\cm^2 \s^{-1}$) and $K_\perp=0$. 
Cosmic rays injected into a localized region  diffuse anisotropically along the
magnetic field lines  and populate a fluxtube marked by magnetic lines
threading the initial injection volume. Due to an excess of cosmic ray pressure
the flux tube becomes underdense and its central part starts to rise against
vertical gravity.  The overall evolution of the fluxtube follows closely the
one described in the thin fluxtube approximation by Hanasz \& Lesch (2000). The
gradient of the cosmic ray pressure accelerates gas, along the direction of
magnetic field, reducing additionally the gas density at the neighborhood of the
injection region. That effect enhances the strength of the buoyancy  force.  

At $t=100$ the cosmic ray populated flux tube forms a rising Parker loop. In
Fig.~\ref{fig6} (c)  the relative density $\Delta \rho/ \rho_0$ is
shown in the $xz$-plane together with the velocity field. The apparent
tube-like cavity in the density distribution results from the local excess of
cosmic rays. The upward gas velocity is a consequence of the buoyancy force.
The rising tube compresses the overlying gas and pushes it toward higher
altitudes. The system perturbed by cosmic rays injected in a localized
spherical region forms a buoyant fluxtube and evolves in a fashion resembling
the development of an undulatory Parker instability mode.

When the diffusion coefficient takes a realistic value $K_\parallel=10^5$, 
which corresponds to $3\cdot 10^{28}\cm^2 \s^{-1}$ the evolution of the system
is remarkably different (see Figs.~\ref{fig7}) and~\ref{fig8}. The distribution
of cosmic rays along the flux tube becomes relatively uniform before the
buoyancy force starts to displace the tube in the vertical direction. The
perturbation provided by the cosmic ray input excites initially (up to $t=100$)
the interchange mode of the Parker instability with a weak contribution of the
undulatory mode. Later on, at $t=150$, the growing contribution of the
undulatory mode becomes apparent.

Due to the more efficient diffusion, cosmic rays fill in initially a larger
volume, compared to the case of lower diffusion coefficient. Therefore the
excess of cosmic ray pressure, and hence the buoyancy force, becomes weaker 
but distributed over a larger volume.  At the fixed time $t=100$ after the
cosmic ray injection the maximum vertical speed is smaller compared to the case
of smaller diffusion coefficient $K_\parallel=10^4$, although later on the
instability accelerates following the emergence of the undulatory mode. The
apparent tendency seems to be opposite to that resulting from the linear
stability analysis by Ryu et al. (2003). However, we point out that the lower
values of the diffusion coefficient are clearly leading to stronger nonlinear
effects, (for instance the vertical speed of the buoyant gas is almost equal to
the sound speed). The large cosmic ray energy remains localized in a more
compact region, therefore the applicability of the linear approximation for the
discussed case of lower parallel diffusion coefficient ($K_\parallel=10^4$) is
questionable. Despite the mentioned differences the late stages of the system
for small and realistic diffusion coefficients remains rather similar.

\section{Summary and conclusions}

In this paper we presented a numerical algorithm for the inclusion of  cosmic
ray dynamics, described by the diffusion-advection equation, into the MHD
code ZEUS-3D. In order to check the presented method we  compared results of
the diffusive passive transport of cosmic rays with analytical solutions
of the diffusion equation. Our method appeared to be numerically  stable in
case of active transport for a wide range of diffusion coefficients, including
the realistic values inferred from the observational data by Strong \&
Moskalenko (1998) for the Milky Way. 

We applied  the presented numerical algorithm to two exemplary simulations of
the excitation of the Parker instability  triggered by cosmic rays injected by
a single SN remnant. The only difference between the input parameters of the
two simulations is the magnitude of the parallel diffusion coefficient. The
simulation corresponding to the realistic value of the parallel diffusion
coefficient $K_\parallel = 3 \times 10^{28} \cm^2 \s^{-1}$, presented in
Figs.~\ref{fig7} and~\ref{fig8}  appeared to develop the Parker instability
mode slower than the one performed for $K_\parallel = 3 \times 10^{27} \cm^2
\s^{-1}$, presented in Fig.~\ref{fig6}. Such a tendency differs from the one
resulting from linear analysis of the Parker instability by Ryu et al. (2003).
We show these two examples of evolution of the system in order to demonstrate
that in some circumstances the finiteness of the diffusion coefficient may lead
to effects which can not be described within the simple linear approximation.
Therefore a verification of all former analytical and numerical results
concerning the nonlinear development of the Parker instability in presence of
cosmic rays is necessary.

The presented work is just a starting point, which focuses on developing the
basic computational techniques. In the next step we plan to combine the cosmic
ray transport in a more realistic application by including the dynamo action of
the cosmic ray component, reconfiguration of the magnetic field by magnetic
reconnection, different random spatial cosmic ray source distributions and
different supernova rates. The future work should also include effects of
cosmic ray losses and extensions of the present algorithm to the energy
dependent cosmic ray transport.

\begin{acknowledgements} This work was supported by the Polish Committee for 
Scientific Research (KBN) through the grant PB  404/P03/2001/20. The presented
work is continuation of a research program realized by MH under the financial
support of {\em Alexander von Humboldt Foundation}. We thank the Laboratory for
Computational Astrophysics, University of Illinois for providing the original
ZEUS-3D MHD code and the referee Dr. A.W. Strong for helpful comments. The
presented computations were done on the Linux cluster HYDRA placed at Torun
Centre for Astronomy.

 \end{acknowledgements}

\bibliographystyle{unsrt}

\end{document}